# Crystal nucleation and near-epitaxial growth in nacre

Ian C. Olson[1], Adam Z. Blonsky[1], Nobumichi Tamura[2], Martin Kunz[2],

and Pupa U.P.A. Gilbert[1,3,*]

[1]Department of Physics, University of Wisconsin–Madison, 1150 University Avenue, Madison, WI 53706, USA.
[2]Advanced Light Source, Lawrence Berkeley National Laboratory, 1 Cyclotron Road, Berkeley, California 94720, USA.
[3]Department of Chemistry, University of Wisconsin–Madison, 1101 University Avenue, Madison, WI 53706, USA.

*Previously publishing as Gelsomina De Stasio. Corresponding author: pupa@physics.wisc.edu

## Abstract

Nacre is a layered, iridescent lining found inside many mollusk shells, with a unique brick-and-mortar periodic structure at the sub-micron scale, and remarkable resistance to fracture. Despite extensive studies, it remains unclear how nacre forms. Here we present 20-nm, 2°-resolution Polarization-dependent Imaging Contrast (PIC) images of shells from 15 mollusk shell species, mapping nacre tablets and their orientation patterns, showing where new crystal orientations appear and how they propagate across organic sheets as nacre grows. In all shells we found stacks of co-oriented aragonite ($CaCO_3$) tablets arranged into vertical columns or staggered diagonally. Only near the nacre-prismatic boundary are disordered crystals nucleated, as spherulitic aragonite. Overgrowing nacre tablet crystals are most frequently co-oriented with the underlying spherulitic aragonite or with another tablet, connected by mineral bridges. Therefore aragonite crystal growth in nacre is epitaxial or near-epitaxial, with abrupt or gradual changes in orientation, with c-axes within 20°. Based on these data, we propose that there is one mineral bridge per tablet, and that "bridge-tilting" is a possible mechanism to introduce small, gradual or abrupt changes in the orientation of crystals within a stack of tablets as nacre grows.

**Introduction**

Nacre is the iridescent inner lining of many mollusk shells. The complex arrangement of nacre tablets inspires biomimetic materials [1-4] yet its formation mechanisms are poorly understood[5-13]. It is well established that interlamellar organic sheets of β-chitin and proteins [5,6,14] are deposited first [15,16], then space is filled by growing aragonite ($CaCO_3$) tablets [15,17]. However, many different models have been proposed for the mechanisms of nacre formation, as summarized in Figure 1. Weiner et al. proposed that organic sheets template aragonite tablet orientation by heteroepitaxy [9,18-20]. "Weiner templates" seemed confirmed as organic molecules induce aragonite growth *in vitro*, rather than calcite [5,21,22]. Another hypothesis by Nudelman et al. had each tablet crystal nucleated independently, by a single, well-defined protein arrangement under each tablet termed "nucleation site" [23]. "Nudelman sites" are highly conserved across species [23]. A third hypothesis by Schäffer et al. had all tablets extending into a myriad of "mineral bridges" through pores in the organic sheets (~100 pores/μm², thus an equal density of "Schäffer bridges" was inferred), an extension of the connected tablet model described by Wada in 1972 [8], thus crystal growth would be homoepitaxial with no new nucleation events at each tablet [10]. Recently Checa et al. showed that the small "Schäffer bridges" in various mollusk species do not connect tablets—they are interrupted by organics, and the crystals across interruptions are not co-oriented [24]. They also showed larger ~200 nm wide mineral bridges near the center of tablets in gastropods and cephalopods, or near a tablet edge in bivalve nacre [24]. The number of "Checa bridges" that exist per tablet could not be addressed by that study. Independent of Checa, Olson et al. proposed a mechanism for nacre growth in which there is one "Checa bridge" at the center of each organic "Nudelman site", thus they represented the "Nudelman sites" as donuts, with a hole at their center [25] (Fig. 1).

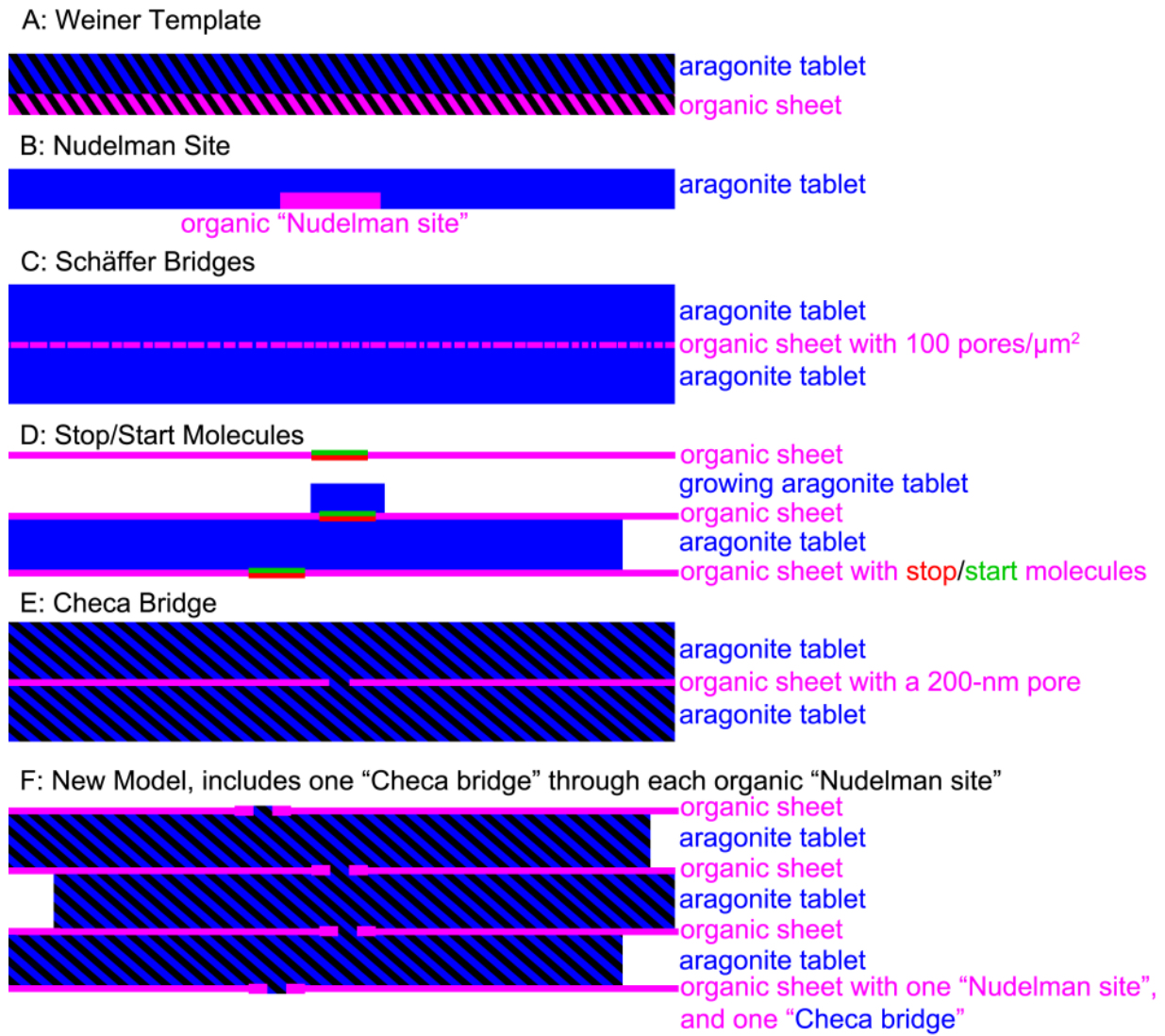

Figure 1. Schematic of all previous and new models proposed for nacre formation mechanisms. (A) In the "Weiner template" model an organic layer of ordered molecules (magenta) imparts the orientation to the overgrowing aragonite crystal tablet (blue) [9,18-20]. The diagonal black dashing represents the orientation, which is templated heteroepitaxially. (B) An organic "Nudelman site" (magenta box) approximately 1-μm in diameter initiates the growth of one aragonite tablet (blue box) [23]. No statements are made about crystal orientations. Two strong and useful statements were introduced by Nudelman et al.: there is a single "Nudelman site" per tablet; the "Nudelman site" is organic. The authors call this "nucleation site" but in light of the present data we prefer not to call this "nucleation", since most frequently each tablet grows homo-epitaxially from the underlying tablet, and does not require a new nucleation event. (C) A myriad of "Schäffer bridges" were supposed to homoepitaxially connect two adjacent tablets (blue) by extending through pores in the organic sheet (magenta) separating two tablets [10]. The "Schäffer bridges" were ruled out by Checa et al. [24]. (D) Metzler et al. [26] hypothesized that one set of organic molecules per tablet is localized on the organic sheets. The chemical nature of these molecules is unidentified, but their function is hypothesized to be as follows: one stop and one start molecule (red, and green, respectively) are in the immediate vicinity of one another and they work in tandem. They are on either side of the same organic sheet, with stop molecule in layer n and start molecule in layer n+1. When growing aragonite crystal n comes in contact with the stop molecule, the start molecule starts the growth of aragonite crystal n+1, co-oriented with crystal n. This model enforces the time sequence of nacre growth, which is key to prevent pores and mechanical failure: no start molecule can nucleate tablet crystals n+1, unless the underlying stop molecule has been reached by crystal n [26]. (E) "Checa bridges" (blue) were found to be 200 nm in diameter, extend through holes in the organic sheet (magenta) separating two subsequent tablets (blue), which are co-oriented (black dashing) [24]. The number of "Checa bridges" present in each tablet could not be determined by the authors. (F) The new model proposed by Gilbert in Olson et al. in which there is one "Checa bridge" poking through one organic "Nudelman site" per nacre tablet [25]. Organic sheets are represented by magenta lines, "Nudelman sites" by thicker magenta donuts with a central hole, and tablets are blue. Black dashing indicates that all three tablets are co-oriented crystals, each growing epitaxially from the underlying one. Notice that, because the organic sheets in both columnar and sheet nacre are formed first, and subsequently growing aragonite crystals fill space, "Weiner templates" must be ruled out. If it were chitin or acidic proteins on the organic layers that template the orientation of newly-nucleated aragonite crystals, these would have to be present and active on the organic sheets at various places and times, even where nacre is not yet mineralized. If this were the case, nucleation could occur before the space underneath is filled, resulting in porous nacre, which is never observed.

Previous work left key questions unanswered: Where and how does nucleation of new aragonite crystals occur? Where and how is the crystal orientation of a tablet transmitted to the overlying tablet? The data here provide new insights into these fundamental aspects of nacre crystal nucleation and growth.

**Results and Discussion**

Using Polarization-dependent Imaging Contrast (PIC) mapping [25-30] at the nano-scale [25,28,29,31], we analyzed shells from 15 species[a] with representative results displayed in Figure 2 and Supporting Information Figure S1. In a PIC-map the gray level corresponds to aragonite or calcite c-axis orientation, and the patterns of orientations of micro-crystalline tablets provide insights into the formation of nacre. Figure 2 shows with unprecedented detail the transition from the calcite prismatic layer to the aragonite nacre layer in *Pf*, *Hi*, and *Np*. In these shells, spherulitic aragonite is revealed by PIC-mapping between the prismatic layer and the first layer of lamellar nacre ($N_o$). Notice that the orientation of tablets at $N_o$ is identical to the orientation of the underlying spherulites, as previously observed in *Np* [32], *Hr* [33,34], and *Pf* [22]. It is evident from the images in Figure 2 that nacre tablet orientation contrast in *Pf*, *Hi*, and *Np* decays with distance from $N_o$. Images of the nacre-prismatic (NP)-boundary in 12 additional species are presented in Supporting Information Figures S1 and S2.

---

[a] These include: the marine bivalves *Atrina rigida* (*Ar*), *Mytilus californianus* (*Mc*), *Mytilus edulis* (*Me*), *Mytilus galloprovincialis* (*Mg*), *Pinctada fucata* (*Pf*), *Pinctada margaritifera* (*Pm*), the marine gastropods *Haliotis discus* (*Hd*), *Haliotis iris* (*Hi*), *Haliotis laevigata* (*HL*), *Haliotis pulcherrima* (*Hp*), *Haliotis rubra* (*Hrb*), *Haliotis rufescens* (*Hrf*), the marine cephalopod *Nautilus pompilius* (*Np*), and the freshwater bivalves *Lasmigona complanata* (*Lc*), and *Pyganodon grandis* (*Pg*). All marine bivalves and gastropods considered in this study have a calcite prismatic layer, while the freshwater bivalves and marine cephalopod have an aragonite prismatic layer.

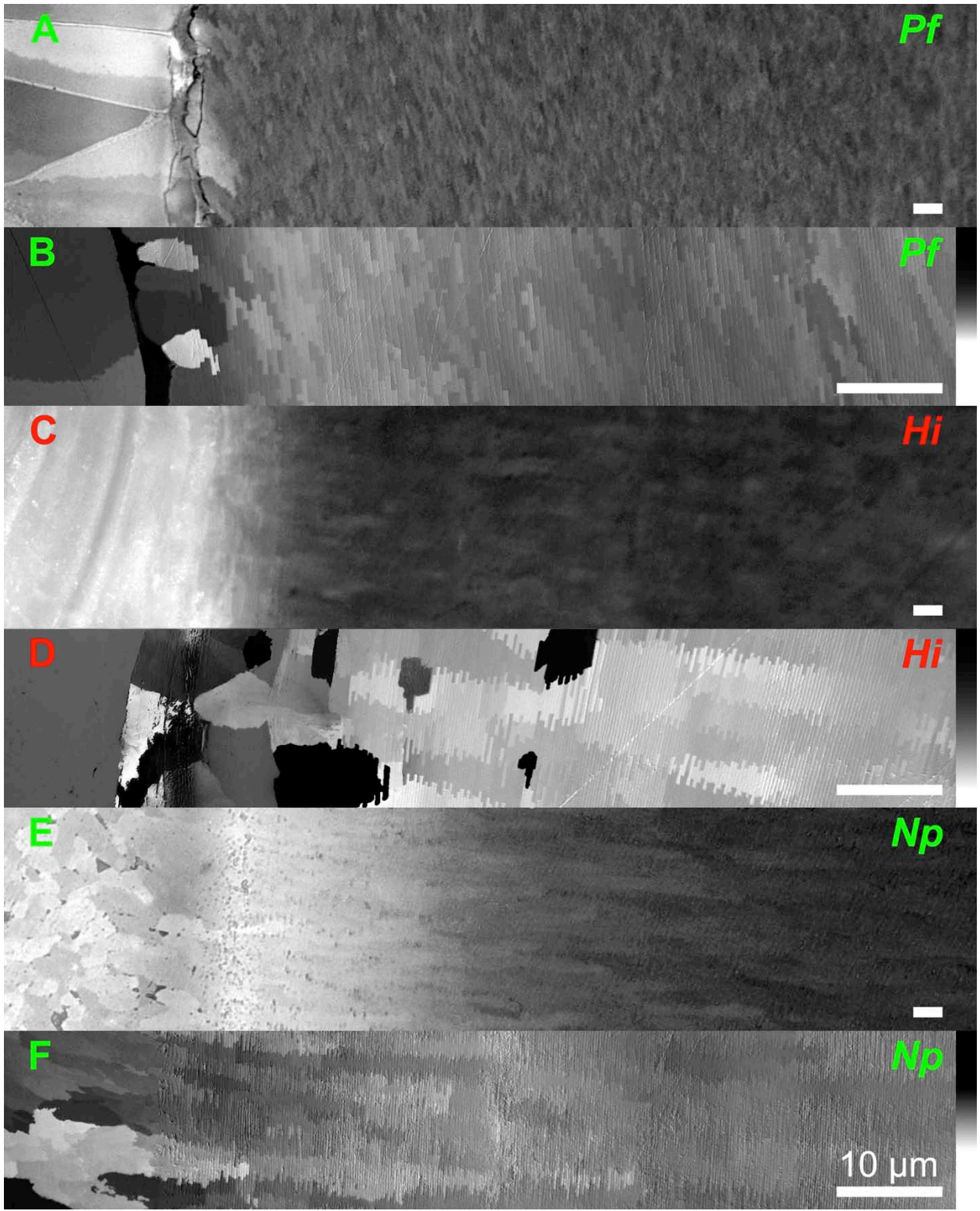

Figure 2. Multi-resolution images of shell cross-sections at the nacre-prismatic (NP)-boundary in the bivalve *Pinctada fucata* (*Pf*), the gastropod *Haliotis iris* (*Hi*) and the cephalopod *Nautilus pompilius* (*Np*). (A,C,E) Polarized, reflected light micrographs illustrating (left) the prismatic calcite (in *Pf*, *Hi*) or aragonite (in *Np*) layer, and (right) the aragonite nacre layer. (B,D,F) PIC-maps of the NP-boundary and nacre in the three species, with 20-nm pixels and 2° angle resolution. The grayscale bars on the right indicate the angle between the aragonite c'-axis of each crystal and the nacre growth direction (horizontal, left-to-right in all panels). The angles ranging from -90° to 90° are displayed in black and white, respectively. The gray level contrast in both types of images is due to different calcite or aragonite c-axis orientations. Contrast decreases with distance from the first nacre layer in these species. Not coincidentally, in all three shells there is disordered spherulitic aragonite at the NP-boundary. All scale bars are 10 μm.

Until recently, nacre in all species was assumed to have co-oriented aragonite crystal c-axes, perpendicular to the shell surface [35]. Many authors observed mis-orientations in nacre c-axes, e. g. 10° [36], but attributed them to the macroscopic shell curvature. Only PIC-mapping as in Figure 2 revealed that this is incorrect at the microscopic scale: immediately adjacent stacks of tablets are dramatically mis-oriented [26].

Striking differences between species in Figure 2, Supporting Information Figures S1, S2, are revealed by PIC-mapping: the morphology of the prismatic layer, the presence or absence of spherulitic aragonite, the arrangement of co-oriented nacre tablets, the degree of crystal mis-orientation, and gradual changes in nacre crystalline order versus distance from $N_o$ [34]. Columnar nacre [14,26,37] from the gastropod (*Hd*, *Hi*, *HL*, *Hp*, *Hrb*, *Hrf*) and cephalopod (*Np*) shells in PIC-maps shows straight columns of co-oriented tablets. Sheet nacre [37,38] in the bivalve shells (*Ar*, *Lc*, *Mc*, *Me*, *Mg*, *Pf*, *Pg*, *Pm*) shows stacks of a few co-oriented tablets staggered diagonally, most clearly visible in *Pf*.

Gilbert et al. showed that, in *Hrf* nacre, aragonite tablet c-axes gradually order along the nacre growth direction as distance from the nacre-prismatic (NP)-boundary increases [34]. This qualitative effect was explained with a model in which faster-growing tablets have their c-axes along the growth direction, and gradually prevail in a competition for space. Here we quantitatively measure the angle spread in each series of PIC-maps, defined as the footprint of the histogram of all c'-axis angles [25], and plot angle-spreads as a function of distance from $N_o$ for all species as shown in Figure 3. The c'-axis is defined as a 2-dimensional projection of the angle between the c-axis and the vertical direction as described in ref. [25]. The c'-axis not the c-axis orientation is measured by PIC-mapping. The 13 shell species in which decay of angle spread is observed have convergence distances 20-400 μm, hence this behavior is typical, and the previously reported 50 μm for *Hrf* [34] is reproduced (Supporting Information Table S1).

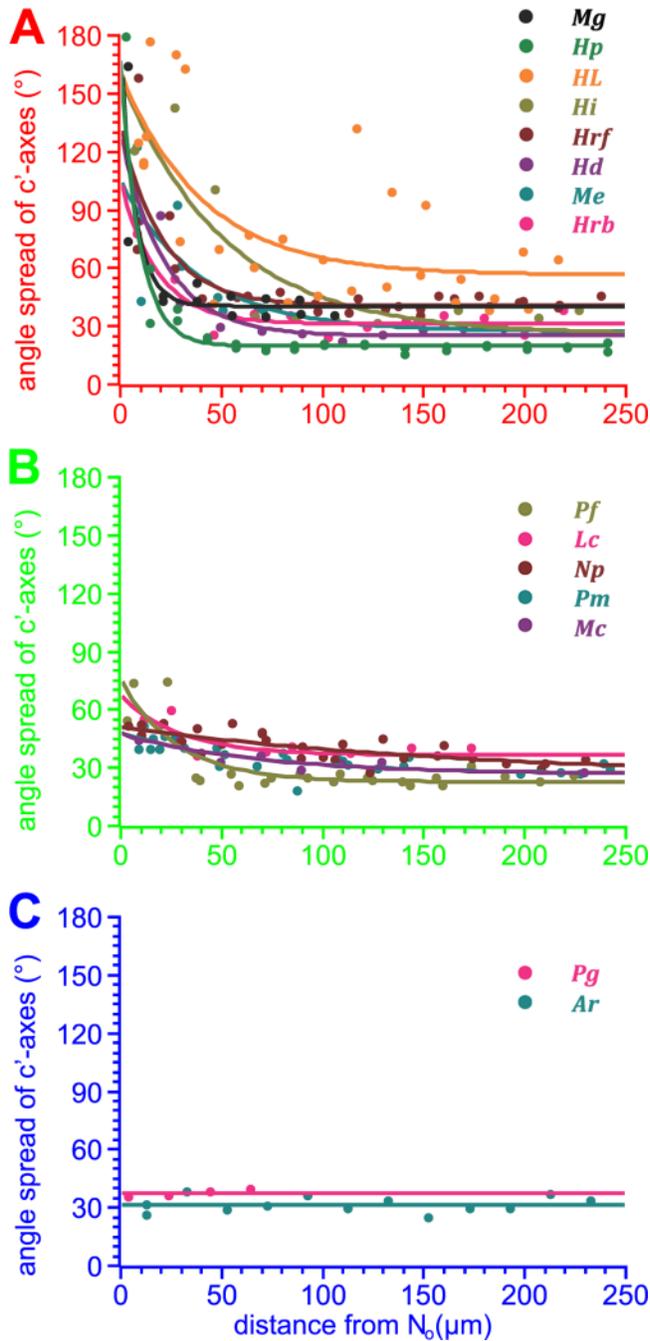

Figure 3. Crystals are nucleated near the NP-boundary, and are most disordered there, then they gradually order in most species. Measurements of angle spread as a function of distance from the first layer of lamellar nacre ($N_o$) in 18 shells from 15 species. Angle spread measurements were obtained from 20 μm × 20 μm PIC-maps of nacre. In a PIC-map, gray level indicates c'-axis orientation. A histogram of all gray levels, therefore, provides the distribution of all angles in a PIC-mapped region of nacre. The angle spread is the full width of this distribution.

The data for each species were fit to an exponential decay (A and B) or to a constant (C). Species names are abbreviated as indicated in the text footnote. All fit parameters are in Supporting Information Table S1. Away from $N_o$ all 15 species have similar angle spreads: 30°±10° (mean ± Std. Dev.). Near $N_o$ the shells differ: most disordered (180° red, A), intermediate (60° green, B), ordered (30° blue, C).

**Aragonite crystal nucleation near the NP-boundary**

The data in Figure 3 show evidence of crystal nucleation near the NP-boundary. In 8 species (c'-axes spread by 180°, Fig. 3A) these are randomly oriented, in 5 others species they are more but not perfectly ordered (60°, Fig. 3B), and in 2 species the crystals have c'axis ordered within 30°, which is as ordered as nacre gets even away from the NP-boundary.

The angle spread of c'-axes in nacre, away from the NP-boundary, is ~30° for all 15 species analyzed here (Fig. 3, Supporting Information Table S1).

These data show beyond any doubt that new nucleations occur near the NP-boundary. Away from the boundary, however, there may or may not be new nucleations. If new tablet nucleations occur in bulk nacre, they must have controlled crystal orientations, and therefore be different from those near the boundary, which exhibit orientational disorder (180°, 60° c'-axis spread) in 13 of the 15 species analyzed.

The formation of ordered nacre at $N_o$ in the remaining 2 species, *Ar* and *Pg*, indicates that these species control crystal orientation at the onset of nacre deposition, possibly by a "Weiner template" mechanism [19] or by inhibiting the growth of crystals with undesirable orientations [39]. This is also the case, although to a lesser extent, in 5 other species (*Pf, Lc, Np, Pm, Mc*), in which the crystal orientation angle spread at $N_o$ is not random but ~60° and decays to ~30° (Supporting Information Table S1). A non-zero amount of crystal orientation control, therefore, must be exerted upon nucleation near the NP-boundary. For the remaining 8 species (*Mg, Hp, HL, Hi, Hrf, Hd, Me, Hrb*), c'-axes angle spread shows that orientations are initially random (~180°). Micro-diffraction results on 3 shells confirm these observations. The spread of c-axes at and away from the NP-boundary is 180°-20° in *Hrf*, 78°-20° in Pm, and in *Ar* the c-axes angle spread is ~20° and constant (Supporting Information Fig. S3). The minor discrepancies between measured c-axis and c'-axis angle spread values likely result from sampling differences or geometry, as explained in Supporting Information Figures S3 and S4.

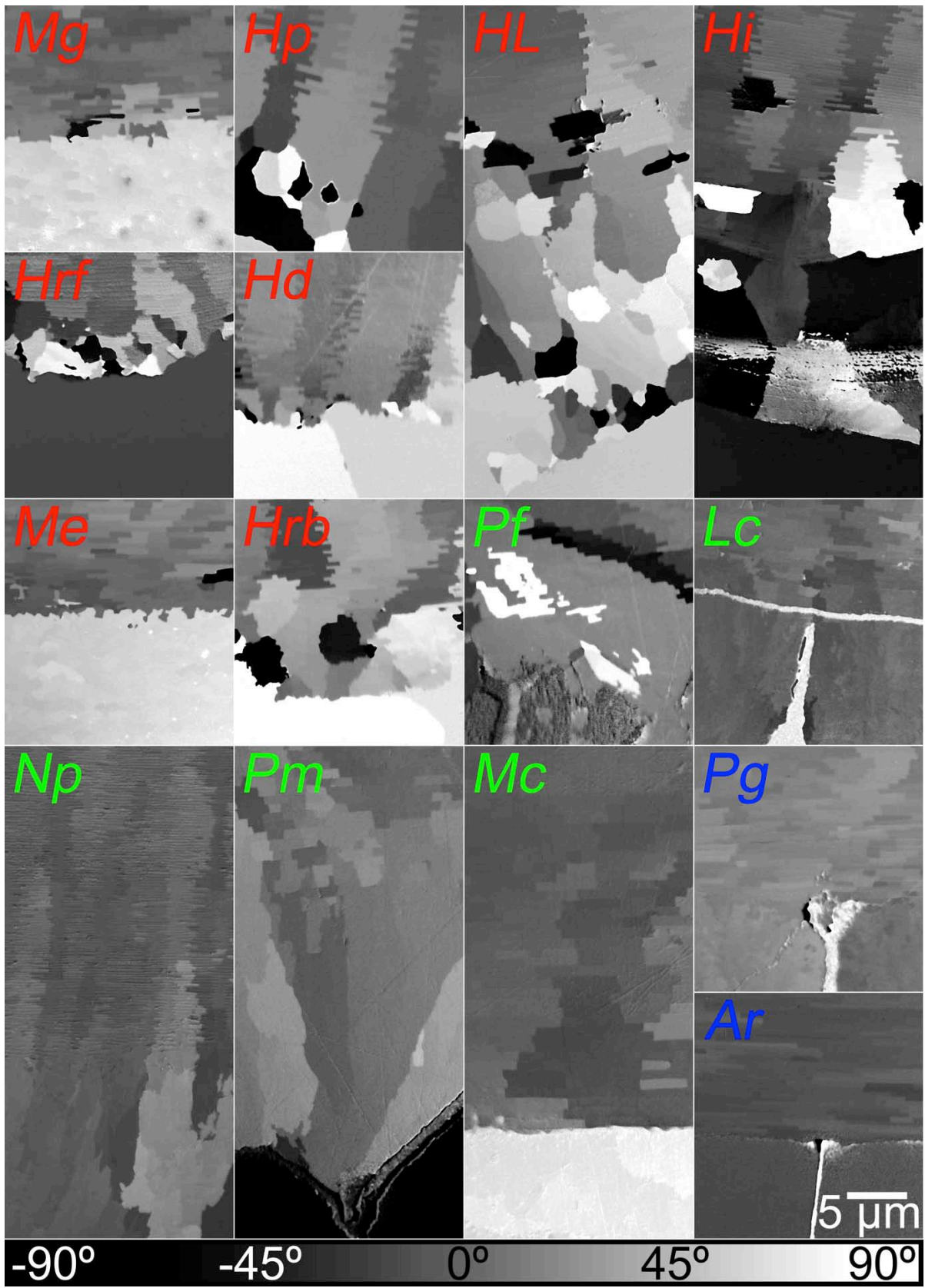

Figure 4. Nacre grows epitaxially from underlying spherulitic aragonite. PIC-maps at the NP-boundary in 15 mollusk species. The c'-axis orientations, displayed as gray levels, show dramatic differences across species. All panels display the prismatic layer at the bottom and nacre at the top. The prismatic layer is calcite in all species, except for *Pg, Lc, Np,* and *Hp*, where it is entirely aragonite. In *Hd* it is mostly calcite, with additional spherulitic aragonite at the outermost surface of the shell (Supporting Information Fig. S5). Species names are abbreviated as indicated in the text footnote, and color-coded as in Figure 3, Supporting Information Figures S1, S2, Table S1. All images share the same 5μm scale bar, and the same gray level bar shown at the bottom.

**Epitaxial crystal growth near the NP-boundary**

Figure 4 presents high-resolution PIC-maps of the NP-boundaries, which exhibit significant differences across the 15 mollusk species. In many of the species in Figure 4, tablets at $N_o$ have the same orientation as the underlying aragonite. This is most strikingly evident in the PIC-maps from all *Haliotis* (*Hp*, *HL, Hi*, *Hrf*, *Hd*, *Hrb*) and *Pinctada* (*Pf*, *Pm*), but also in some of the *Mytilus* (*Mg*, *Me* but not *Mc*), *Lc*, *Np,* and *Pg* shells. Remarkably, in 13 of the 15 species the angle-spread decay behavior observed in Figure 3 is correlated with the existence of disordered spherulitic aragonite at the NP-boundary (Fig. 4, Supporting Information Table S1). The observation of c'-axis orientation shared between nacre tablets and the underlying aragonite spherulites (Figs. 2, 4, Supporting Information Figs. S1, S2) suggests that aragonite crystals grow homoepitaxially, possibly across pores in nacre organic sheets, as shown in Figure 1F. Further support for homoepitaxy was found in an unusual location of the *Np* shell, where striking transitions occur from nacre to spherulitic aragonite, to nacre again, in which the crystal orientations are consistently homoepitaxial across nacre interlamellar organic sheets, as shown in Supporting Information Figure S6. Holes in the sheets (Fig. 1F) provide the simplest explanation for the observed continuous, uninterrupted, homoepitaxial crystal growth of aragonite. More complex mechanisms such as that described by Metzler et al. [26] (Fig. 1D) are also plausible, but do not explain the "Checa bridges" so clearly and reproducibly shown by Checa et al. [24]. We recently proposed that "Checa bridges" through such holes provide the topological link between tablets [25]. The data presented here do not conclusively demonstrate but appear to support that hypothesis: most frequently each stack of tablets is a single crystal with c-axis orientation alignment of 2° or better. The same extent of alignment is expected for the a- and b-axes.

How many "Checa bridges" are there per tablet? Independent lines of evidence suggest that there is a single bridge per tablet: the Voronoi construction of tablets in each nacre layer [34,38] can only be formed if there is a single seed for the growth of each tablet; the gradient of orientation in Figure 5, and its possible explanation by "bridge-tilting" in Supporting Information Figure S7, invoke a single growth center per tablet. If there is one "Nudelman site" and one "Checa bridge" per tablet, these must be the same entity. This is why we propose here that one "Checa bridge" extends through each "Nudelman site". We stress that this is a logical deduction, not an experimental observation, as the PIC-maps presented here do not directly reveal either "Checa bridges" or "Nudelman sites".

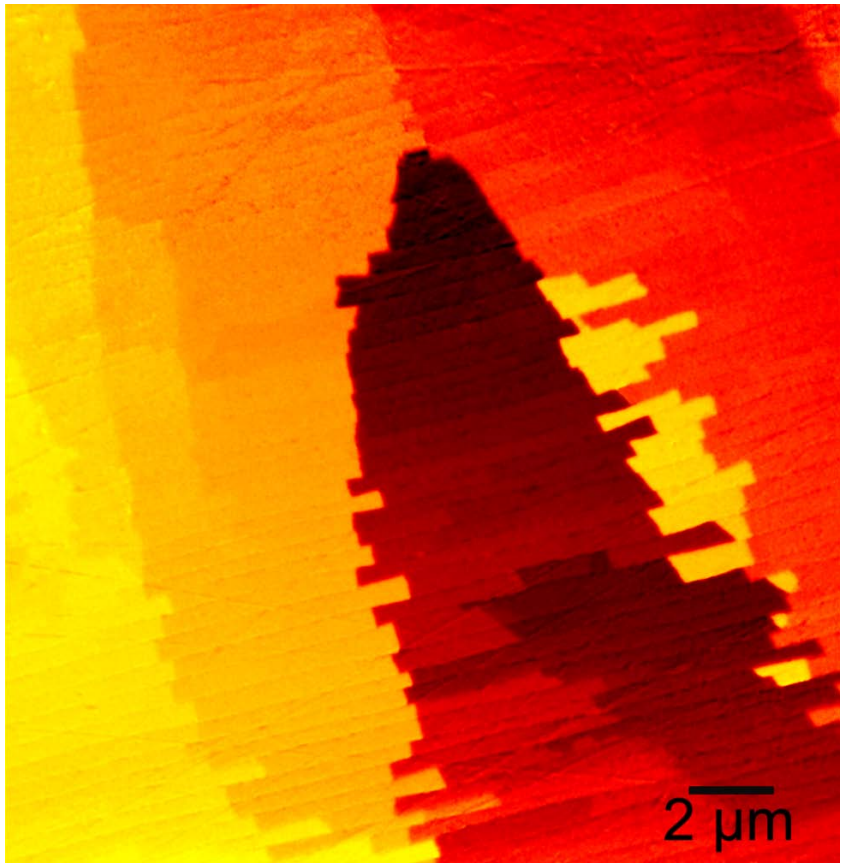

Figure 5. False-colored PIC-map of nacre in *Haliotis laevigata*. Stacks of co-oriented tablets are homogeneously colored, whereas the central dark stack exhibits a gradient of colors. A possible interpretation of this observation is as follows: a single "Checa bridge" per tablet, slightly tilted in orientation as it grows, seeds the orientation of the overlying tablet. Similar tilting in the same direction is repeated at the next bridge and tablet. A new nucleation event, as occurs near the NP-boundary, would have c'-axis angle spreads in the range between ±30° and ±90° (Figure 3). "Bridge-tilting" is a more plausible explanation for the gradient of orientations seen here. Additional data suggesting bridge tilting are presented in Supporting Information Figures S8 and S9. This hypothetical mechanism is schematically described in Supporting Information Figure S7.

**Near-epitaxial crystal growth away from the NP-boundary**

To further explore the possibility that nacre grows epitaxially through a single "Checa bridge" per tablet, as proposed here, we searched for new crystal orientations and nucleations away from the NP-boundary. High resolution PIC-maps of nacre away from the NP-boundary are presented in Supporting Information Figures S8 and S9. The PIC-maps are not as homogenous in orientation as one might have expected, but they may still rule out new nucleations. From the PIC-maps in Figure 4 and the µXRD data in Supporting Information Figure S3 one would expect new crystals to have c'-axis angle spreads in the 60°-180° range, not the 30° range observed in bulk nacre (Fig. 3, Supporting Information Figs. S3, S4, Table S1). We propose "bridge tilting" as a hypothetical model to explain abrupt or gradual orientation changes in the same stack of tablets, as observed in Supporting Information Figures S8 and S9. The alternative mechanism, new nucleation events, cannot by themselves explain the alignment of new c'-axis orientations within 30°. New crystal nucleation on a "Weiner template" (Fig. 1A) could, in principle, orient newly nucleated aragonite crystals, but would necessarily imply uncontrolled nucleation time and location away from the growing nacre front [15,17], thus generating discontinuous, non-space-filling nacre, which is never observed (Fig. 1 bottom of caption). Organic bridges such as the stop/start hypothetical molecules described by Metzler et al. [26] (Fig. 1D) could also explain the observed small changes, but to date there is no evidence for the existence of these hypothetical molecules. Because we have hard evidence from Checa et al. [24] for "Checa bridges", we believe that the most plausible explanation is "bridge tilting" as described in detail in Supporting Information Fig. S7. But we stress that this is, at this time, a hypothesis and we do not have direct evidence for bridge tilting. Only high-resolution transmission electron microscopy studies of nacre could provide evidence to support or rule out "bridge tilting".

**Conclusions**

Clearly both nucleation and growth of crystals are highly regulated in mollusks. Because seawater, and presumably the extrapallial fluid from which nacre is formed [40], is super-saturated with respect to aragonite, nucleation at random times, positions, and orientations must be actively inhibited. Sequential, connected, near-epitaxial aragonite crystal growth is enabled and regulated [22,35] everywhere in nacre, near and away from the NP-boundary, whereas nucleation is enabled near the boundary and may or may not be enabled in bulk nacre.

In all shells we found nearly co-oriented tablets stacked into columns or staggered diagonally, and in many cases these are co-oriented with underlying tablets or spherulitic aragonite. In conclusion, the data here provide evidence for near-epitaxial growth of nacre, and we propose that one "Checa bridge" in one "Nudelman site" per tablet connects all tablets in each stack, with c'-axes and c-axes angles spread by 30° and 20°, respectively.

It is unclear at present whether greater co-orientation of nacre tablets provides an evolutionary advantage to the mollusk. It is tempting to conclude that such advantage must exist, because the shells that do not start with ordered nacre rapidly achieve such order within the first 20-400 μm.

Although the pattern of crystals in nacre, their orientation, and their formation mechanism are further understood in this work, the origin of the regularity of the crystal layer thickness remains a mystery. This is ironic, if one considers that iridescence is the most conspicuous characteristic of nacre.

**Brief Methods**

The shells of fifteen mollusk species were cut, embedded in epoxy (either EpoThin, Buehler, IL, or EpoFix, Electron Microscopy Sciences, PA), polished perpendicularly to the nacre layers with decreasing size alumina grit down to 50 nm (MasterPrep, Buehler, IL), and coated with Pt using a sputter coater (208HR, Cressington, UK). Samples were analyzed with PEEM-3 on beamline 11.0.1 and synchrotron Laue micro-X-ray diffraction on beamline 12.3.2 at the Advanced Light Source at Lawrence Berkeley National Laboratory in Berkeley, CA. Detailed Materials and Methods are provided in the Supplementary Information.

*PIC-mapping with PEEM*

Nineteen images were collected at the same 290.3 eV photon energy, and a sample voltage of -15 kV, while the linear X-ray polarization vector was rotated between 0°-90° in 5° increments. Each pixel of these stacks of 19 images, therefore, contained a polarization-dependence curve which was fit to the function y = a + b cos(EPU° + c). The analysis was repeated for all $10^6$ pixels in each stack of 20 μm × 20 μm images, with 20-nm pixel size, producing a PIC-map image in which the fit parameters c were displayed as quantitative gray levels. Angle spread measurements were taken as the "footprint" of the distribution of angles in a PIC-map.

*µXRD analysis*

Samples were illuminated with "pink beam" x-rays of photon energies ranging between 5<hν<22 keV. X-ray microdiffraction patterns were obtained using a Pilatus 1M X-ray detector. Laue X-ray microdiffraction patterns were indexed using the XMAS software [41] (X-ray Microdiffraction Analysis). Indexing provides the full 3-dimensional orientation matrix for each crystal, allowing for the mapping of the distribution of orientations of aragonite crystallites in the sample.

**Acknowledgements**

We thank Sabine Gross for providing the *Lc* and *Pg* shells, and Lisie Kitchel for the identification of these species. We thank Steve Weiner for the *Hp* shell, collected by Heinz Lowenstam, and identified by David Lindberg. We thank ALS beamline scientists Andreas Scholl and Anthony Young for their technical support during the PEEM-3 experiments. We thank Fabio Nudelman, Steve Weiner, and Amir Berman for reading the manuscript and suggesting improvements, and Lia Addadi for discussion. This work was supported by NSF award DMR-1105167 and DOE Award DE-FG02-07ER15899 to PUPAG. The experiments were performed at the Berkeley Advanced Light Source, supported by DOE under contract DE-AC02-05CH11231.

**Supporting Information Available**. Ten additional figures and detailed methods.

**TOC graphic**

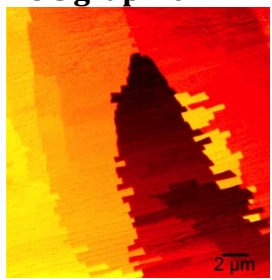

# Supporting Information
# for

## Crystal nucleation and near-epitaxial growth in nacre

Ian C. Olson[1], Adam Z. Blonsky[1], Nobumichi Tamura[2], Martin Kunz[2],

and Pupa U.P.A. Gilbert[1,3,*]

[1]Department of Physics, University of Wisconsin–Madison, 1150 University Avenue, Madison, WI 53706, USA.
[2]Advanced Light Source, Lawrence Berkeley National Laboratory, 1 Cyclotron Road, Berkeley, California 94720, USA.
[3]Department of Chemistry, University of Wisconsin–Madison, 1101 University Avenue, Madison, WI 53706, USA.

*Previously publishing as Gelsomina De Stasio. Corresponding author: pupa@physics.wisc.edu

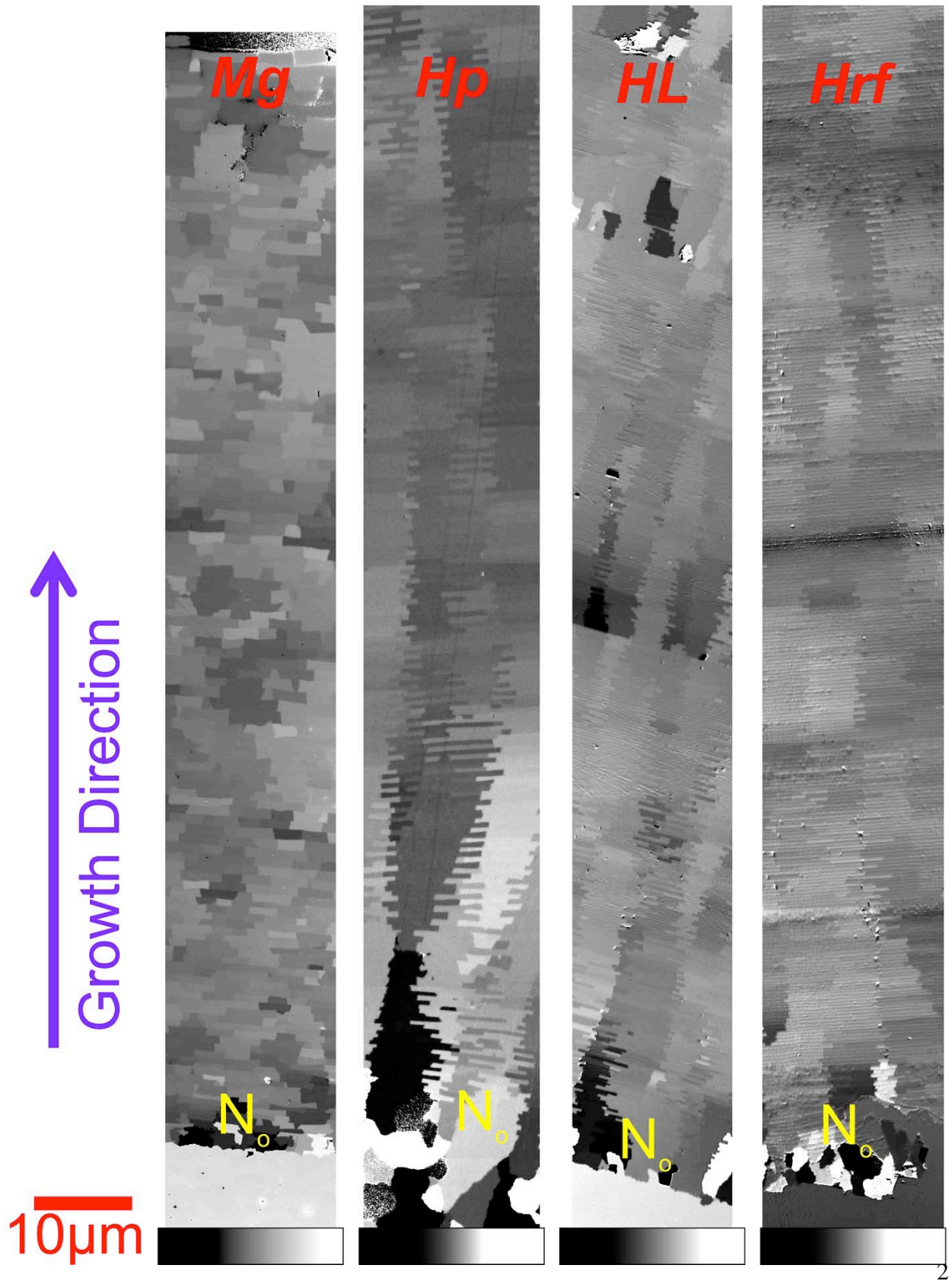

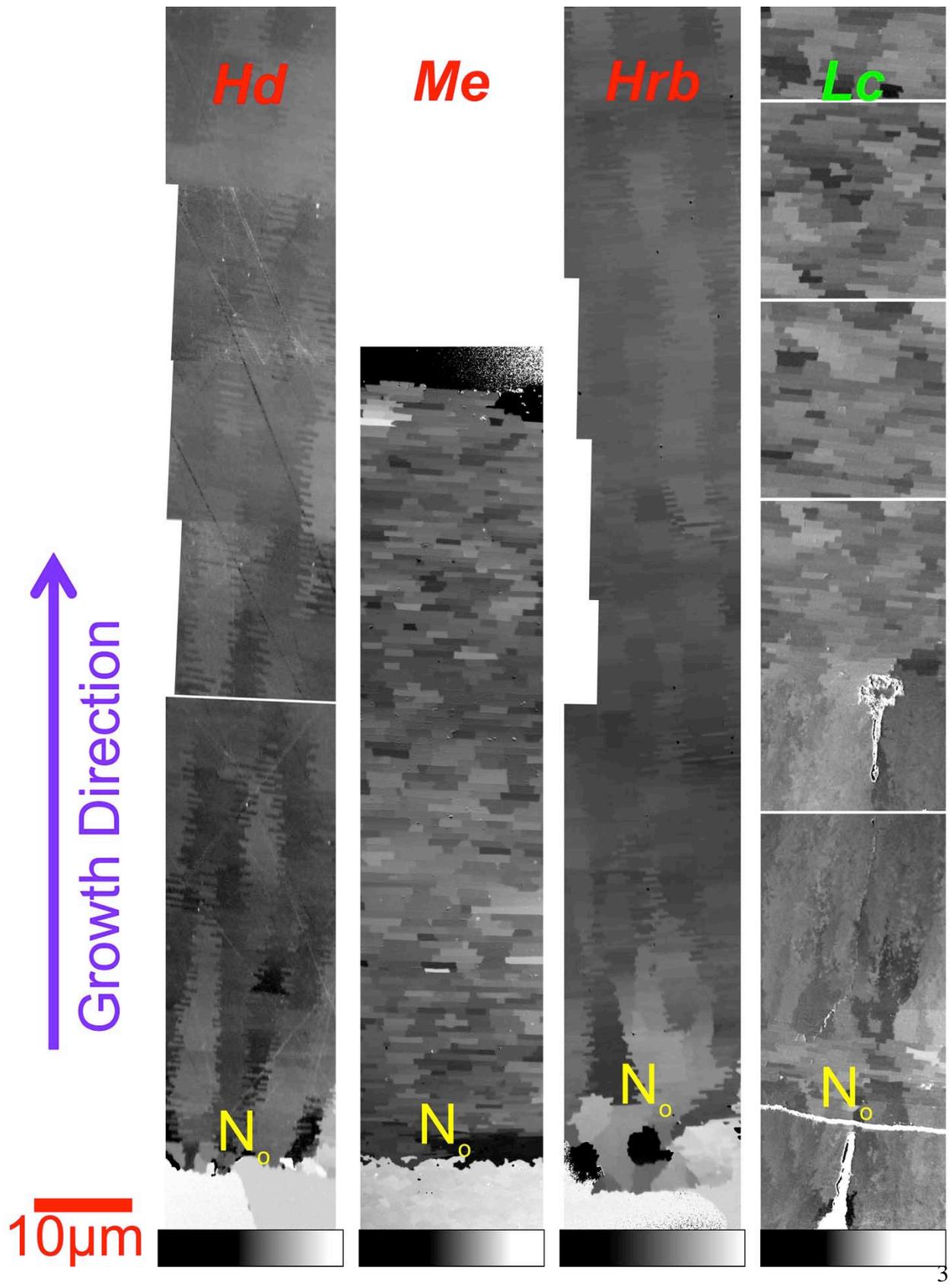

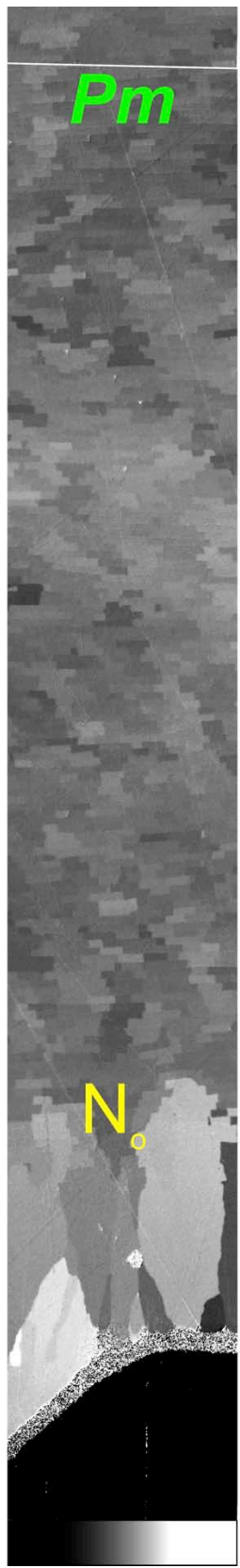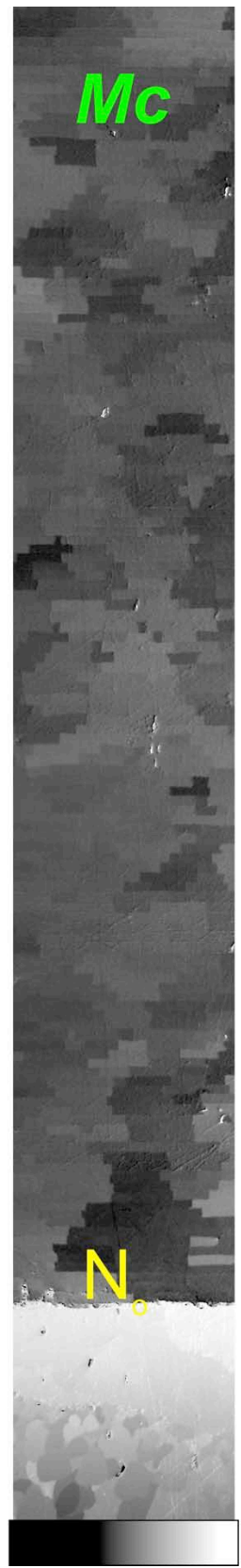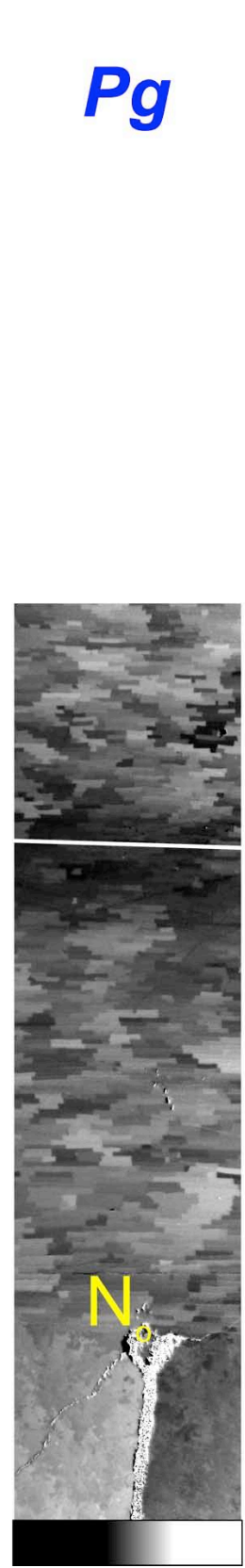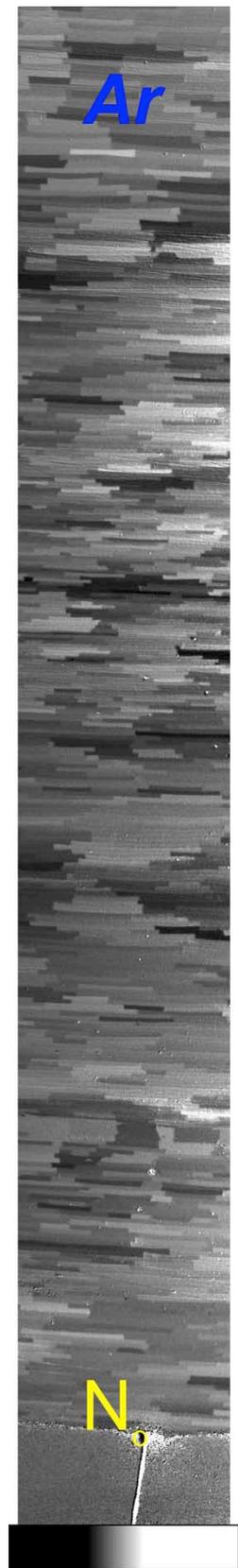

Supporting Information Figure S1. PIC-maps from shell cross-sections of 12 mollusk species. Each series begins at the NP-boundary with the first lamellar nacre layer labeled as $N_o$, and extends 60-110 μm into the nacre layer. The growth direction, that is, the direction in which the macroscopic thickness of nacre increases, is indicated by a purple arrow, pointing toward the more recently deposited nacre at the inner surface of the shell. The grayscale bar at the bottom of each series indicates the angle, ranging from -90° to 90°, displayed in black and white, respectively, between the aragonite c'-axis of each crystal and the nacre growth direction. Each series and grayscale bar set was independently leveled to enhance contrast. Species name abbreviations, as defined in the main text, are: *Mytilus galloprovincialis* (*Mg*), *Haliotis pulcherrima* (*Hp*), *Haliotis laevigata* (*HL*), *Haliotis iris* (*Hi*), *Haliotis rufescens* (*Hrf*), *Haliotis discus* (*Hd*), *Mytilus edulis* (*Me*), *Haliotis rubra* (*Hrb*), *Pinctada fucata* (*Pf*), *Lasmigona complanata* (*Lc*), *Nautilus pompilius* (*Np*), *Pinctada margaritifera* (*Pm*), *Mytilus californianus* (*Mc*), *Pyganodon grandis* (*Pg*), *Atrina rigida* (*Ar*). We note here and in Figures 2 and 4 that *Pm* and *Pf* have a thick organic layer separating calcite from spherulitic aragonite; *Mg*, *HL*, *Me*, *Mc*, *Hd*, and *Hrb* have their calcite prismatic layer in direct contact with aragonite with no evidence for an organic envelop separating the two minerals; there is also no evidence for any co-orientation of the calcite and aragonite c-axes in immediately abutting crystals. The two latter observations are also verified in another region of *Hd*, as described in Supporting Information Figure S5. *Hp*, *Lc*, *Pg*, and *Np* have an aragonitic prismatic layer, and again no obvious organic layer separating prismatic aragonite from lamellar nacre (the feature at the interface in *Lc* is a crack, as shown by SEM imaging).

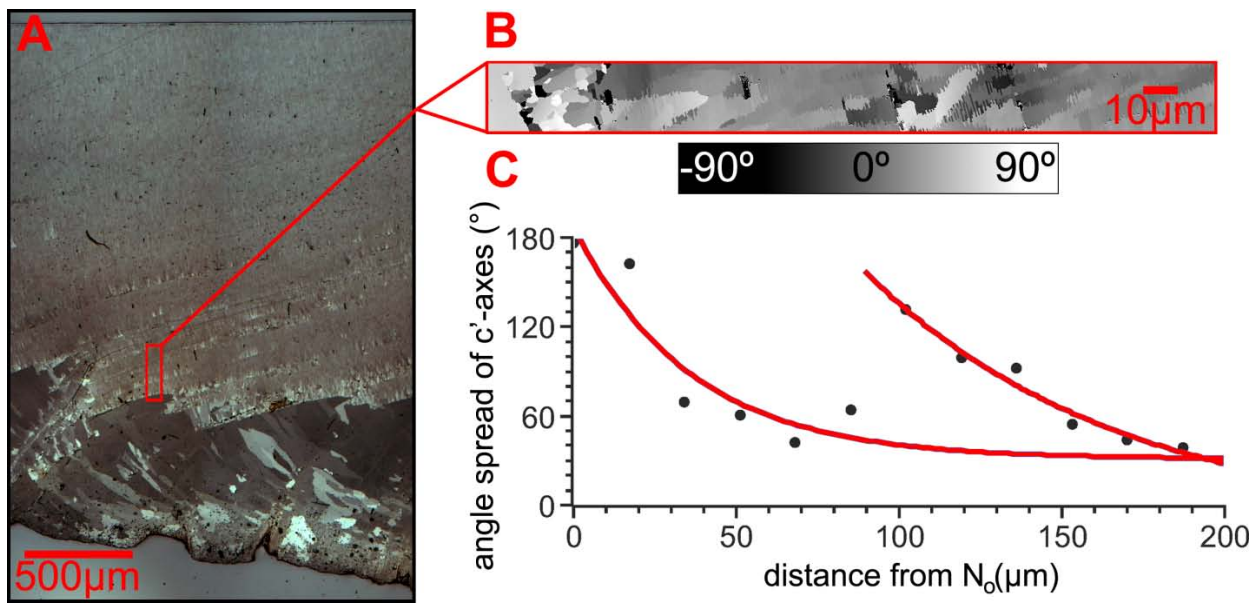

Supporting Information Figure S2. Multiresolution images illustrating disorder in the nacre of *Haliotis laevigata* (*HL*) resulting from bands of highly-disordered blocky aragonite. (A) VLM image of the nacre (top) and prismatic (bottom) layers in *HL*. Note the high-contrast bands in the nacre layer. (B) PIC-maps of the NP-boundary and nacre in *HL* demonstrating the appearance of highly disordered aragonite. The grayscale bar indicates the angle between the aragonite c'-axis of each crystal and the nacre growth direction, left-to-right in the PIC-map in (B). (C) Plot of the angle spread of the c'-axis in *HL* from the regions in (B).

**Supporting Information Table S1**

| species | angle spread at N₀ (°) | convergence distance d (μm) | steady-state c'-axis angle spread y₀ (°) | spherulitic aragonite present at the boundary? |
|---|---|---|---|---|
| *Mg* | 170±70 | 17.0±0.3 | 40±8 | Yes |
| *Hp* | 166±5 | 21.87±0.02 | 20±1 | Yes |
| *HL* | 160±30 | 93.51±0.03 | 60±10 | Yes |
| *Hi* | 160±30 | 117.63±0.02 | 30±10 | Yes |
| *Hrf* | 130±20 | 47.82±0.04 | 41±4 | Yes |
| *Hd* | 130±20 | 49.61±0.03 | 25±4 | Yes |
| *Me* | 110±70 | 89.7±0.1 | 30±50 | Yes |
| *Hrb* | 110±20 | 41.01±0.04 | 31±2 | Yes |
| *Pf* | 76±9 | 63.10±0.03 | 23±3 | Yes |
| *Lc* | 70±10 | 66.21±0.06 | 37±3 | Yes |
| *Np* | 51±8 | 393.90±0.01 | 25±6 | Yes |
| *Pm* | 48±4 | 158.55±0.01 | 27±2 | Yes |
| *Mc* | 48±5 | 156.37±0.02 | 27±3 | No |
| *Pg* | 37.3±0.8 | 0 | 37±1 | Yes |
| *Ar* | 31±1 | 0 | 31±1 | No |

Supporting Information Table S1. All fit parameters for the fits in Figure 3. The experimental data from each shell were fit to either an exponential decay of the form $y = y_0 + Ae^{-x/B}$ or a constant $y = y_0$. In these fit equations, $y_0$ represents the steady-state value of the angle spread, $y_0+A$ represents the angle spread at N₀, and $d$ is the convergence distance at which the angle spread has decayed by 90%, that is, to a value $y = y_0 + A/10$. The convergence distance is also $d = (-\ln 0.1)B$. The convergence distance is zero for species fit to a constant. The fit parameters are colored according to their initial high (Fig. 3A) or low (Fig. 3B) angle spread at N₀, which then decays with distance from N₀, or constant (Fig. 3C) angle spread. Interestingly, there is no correlation (R=0.28) between angle spread at N₀ and convergence distance $d$. *Ar* and *Pg* do not show an exponential decay of angle spread in Figure 3. We note that angle-spread decay occurs almost exclusively in shells with disordered spherulitic aragonite at the NP-boundary. *Mc* and *Pg* are the only exceptions: *Mc* has no spherulitic aragonite but its angle spread decays from ~50° to ~30°, whereas *Pg* has spherulitic aragonite but its angle spread is constant at ~37° from the onset of nacre formation.

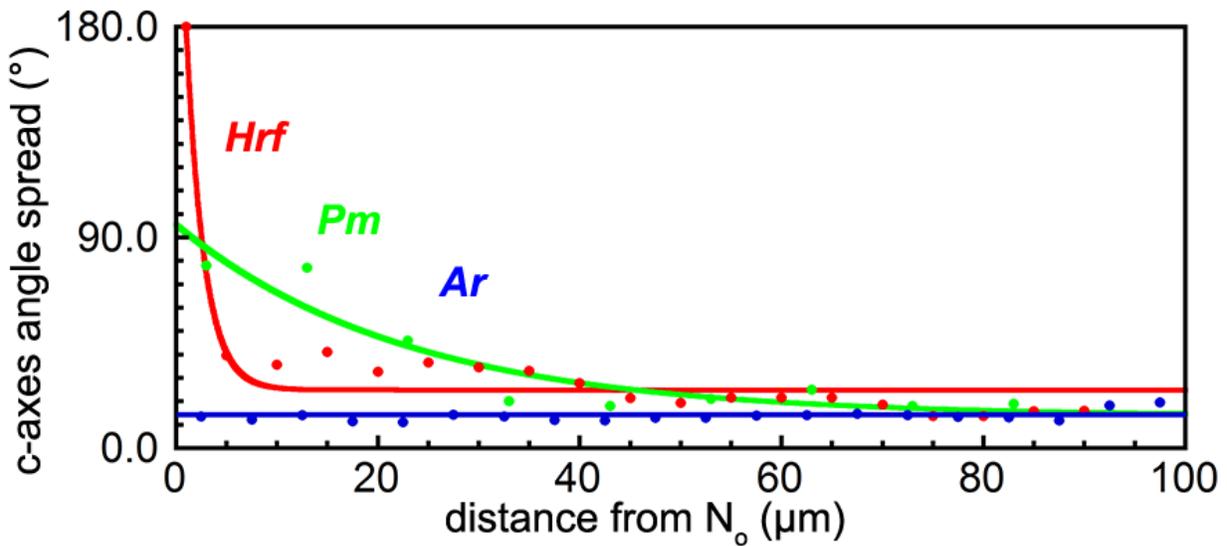

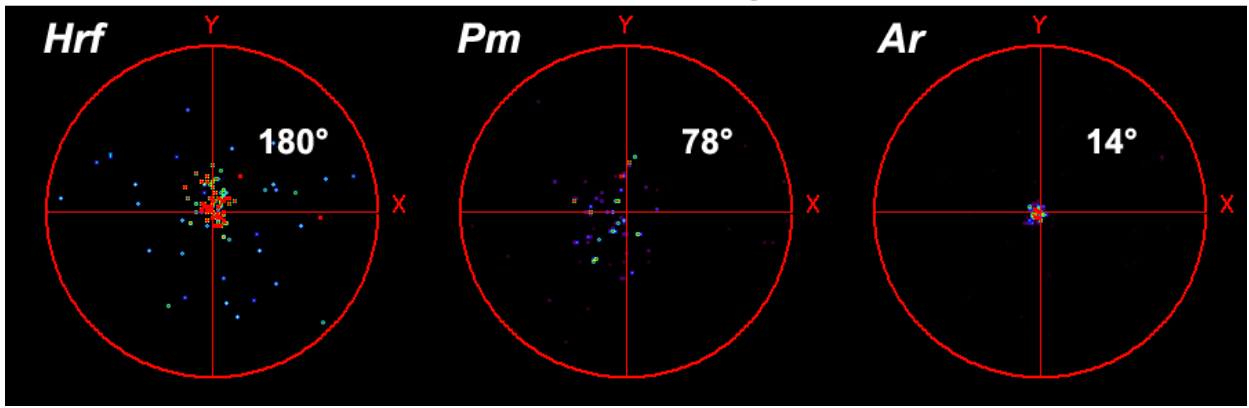

Supporting Information Figure S3. Synchrotron Laue micro-X-ray diffraction (µXRD) results. Plots of nacre c-axis angle spread vs distance from $N_o$ (top) and aragonite c-axis pole figures in nacre at $N_o$ zoomed to an angular range of ±90° (bottom). The decay of angle spread with distance from $N_o$ in *Hrf* and *Pm*, and the constant angle spread observed in *Ar*, corroborate the results for these species in Figure 3. The µXRD angle spreads represent the maximal angular spread of the orientation of the c-axis for all indexed crystallites (sets of co-oriented tablets) obtained from a two-dimensional diffraction scan. These data probe a larger area and volume and therefore many more crystallites than a PIC-map. They measure the c-axis orientations in three-dimensions, rather than their two-dimensional projections, the c'-axes. The c-axis angle spreads measured by µXRD for *Hrf* and *Pm* at $N_o$ are larger than the c'-axis angle spreads measured with PIC-mapping. This must be due to sampling many more crystal orientations in the bulk-sensitive µXRD data, compared to the surface sensitive PIC-mapping. The µXRD data give a more accurate representation of real 3-dimensional angle spread, although with much lower spatial resolution. The angle spread at steady-state, that is, away from the NP-boundary is 20° in all three species analyzed by µXRD. This is smaller than the 30° c'-axis angle spread shown in Figure 3. Supporting Information Figure S4 shows how, for all shells away from the NP-boundary, a 20° angle in 3D may appear as 30° in a 2D projection.

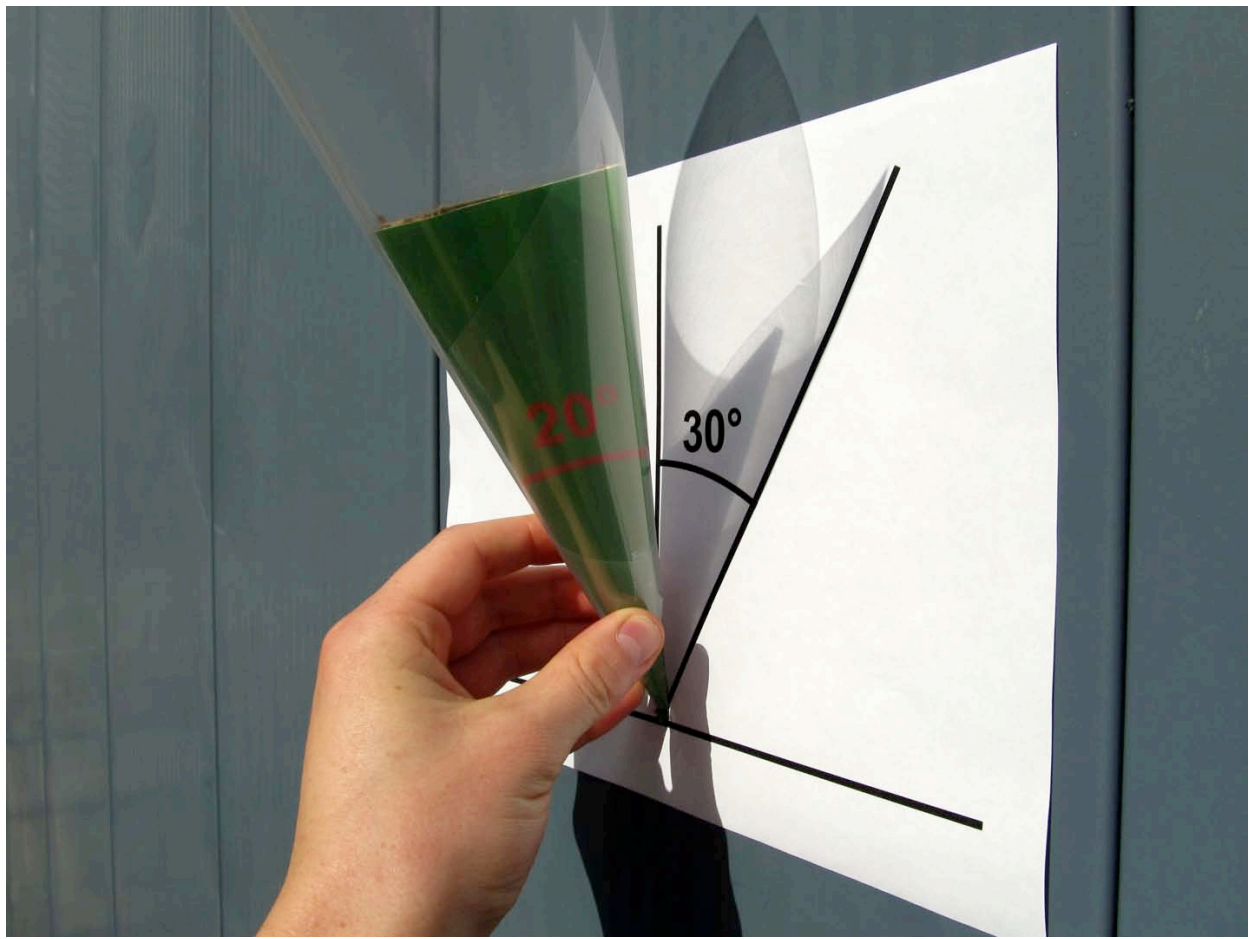

Supporting Information Figure S4. A cone of aperture 20° illuminated by sunlight casts a shadow on a two-dimensional plane that is 30°. This simple demonstration explains quantitatively how the steady-state angle spread measured by PIC-mapping is overestimating (30°) the real angle spread measured by µXRD (20°). Notice that when projecting a 20°-aperture three-dimensional cone onto a two-dimensional plane, any angle can be observed between 20° and 180°, but not <20°. This statement is true for shadows with perpendicular illumination or for PIC-mapping, as they both are two-dimensional projections of a three-dimensional cone.

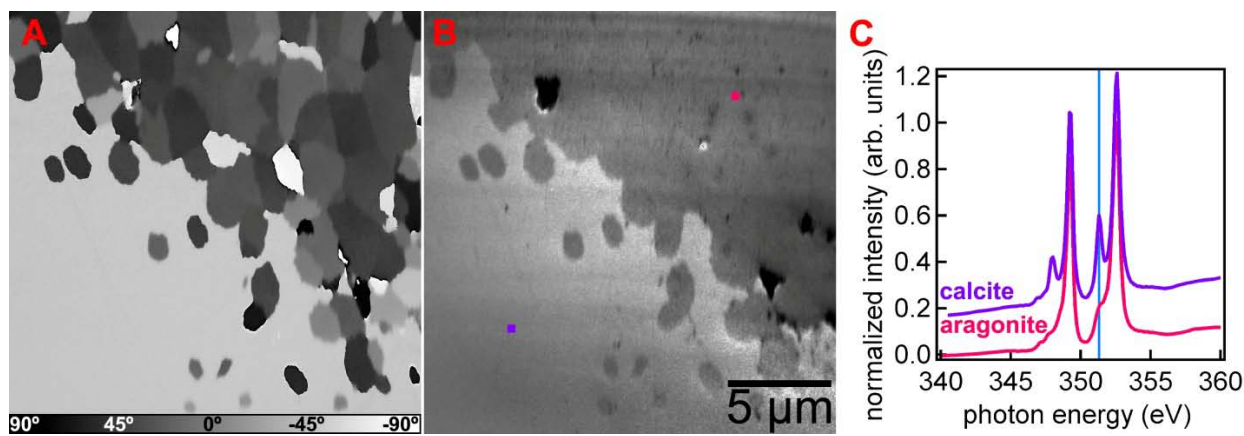

Supporting Information Figure S5. Interface between spherulitic aragonite in the outer layer and prismatic calcite in the shell of *Haliotis discus* (*Hd*). Such outer spherulitic aragonite shell structure is reminiscent of that observed at the outer surface of *Nautilus*, and to the best of our knowledge it had never been reported for any of the *Haliotis* species. (A) PIC-map, in which the orientation of the aragonite or calcite crystal c'-axis with respect to the vertical direction in the plane of the image is indicated by the grayscale bar at the bottom. The image has been leveled to enhance contrast. Notice the disordered spherulitic aragonite on top right, and a large single crystal of calcite at the bottom left of the image. (B) Calcium map, illustrating the spatial distribution of calcite (bright) and aragonite (dark) in the same region as A, obtained by the ratio of photoelectron emission micrographs taken on- and off-peak (at 351.3 eV and 340 eV, respectively). Calcite appears bright because there is a crystal-field peak at 351.3 eV, which is much more intense in calcite than in aragonite, as shown by the spectra and the vertical turquoise line in C. (C) X-ray absorption near-edge structure (XANES) spectra of calcite and aragonite averaged from the 400-nm square regions in B, correspondingly colored. The crystal-field peak at 351.3 eV is indicated by a vertical turquoise line. The presence of aragonite and calcite in direct contact with one another was not previously reported, but is observed in several of the species in Figures 4 and Supporting Information Figure S1 (*Mg*, *Me*, *Mc*, *Hd*, and *Hrb*). Similarly, the images in this figure show regions of highly disordered spherulitic aragonite interspersed within a single crystal of calcite, a region with seemingly uncontrolled aragonite nucleation. In these species, adjacent calcite and aragonite crystals show dramatically different c'-axis orientations indicating a lack of interaction between these abutting crystals during shell formation.

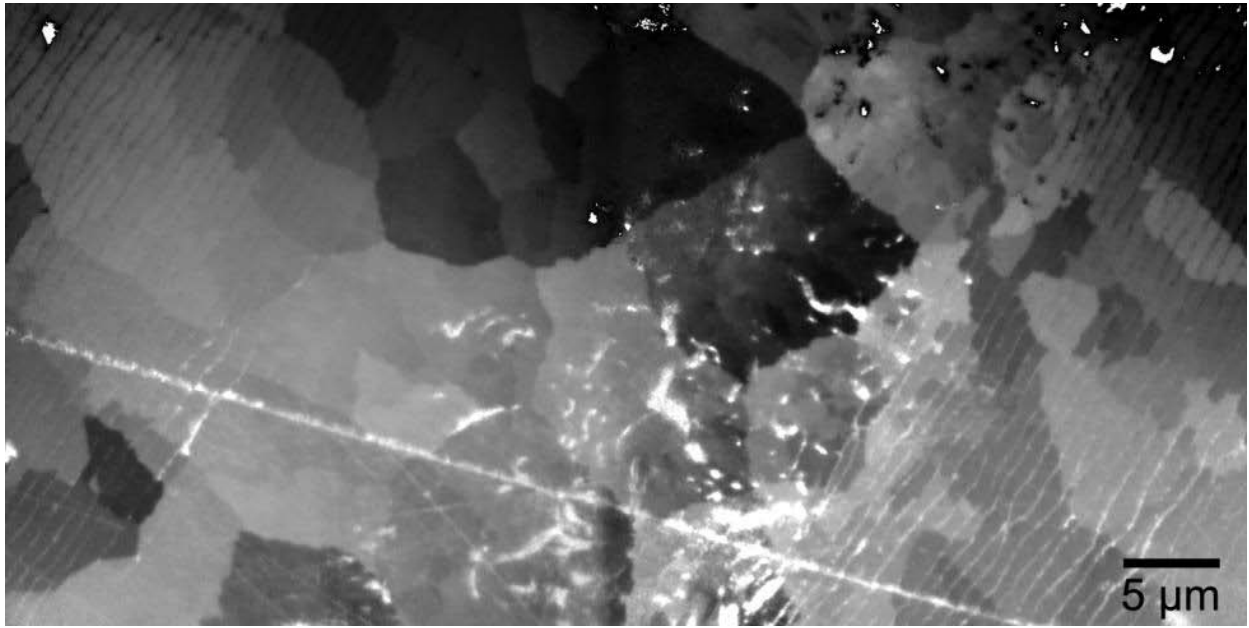

Supporting Information Figure S6. PIC-map of a cross-section of the umbilical callus in *Nautilus pompilius*. The regions with interlamellar organic sheets on the left and the right are nacre, and regions without organic sheets at the center are spherulitic aragonite. Note that the nacre orientation appears to be inherited from adjacent spherulitic aragonite, and *vice versa*. This image demonstrates that nucleation is not inhibited in this part of the shell, hence the appearance of randomly-oriented layered or spherulitic aragonite. This observation strongly suggests that nacre layering and nucleation-inhibition are independent processes. It also strongly supports the conclusion that crystal growth is epitaxial across nacre interlamellar organic sheets, as the orientation of crystals is completely unaffected by the appearance or disappearance of the sheets.

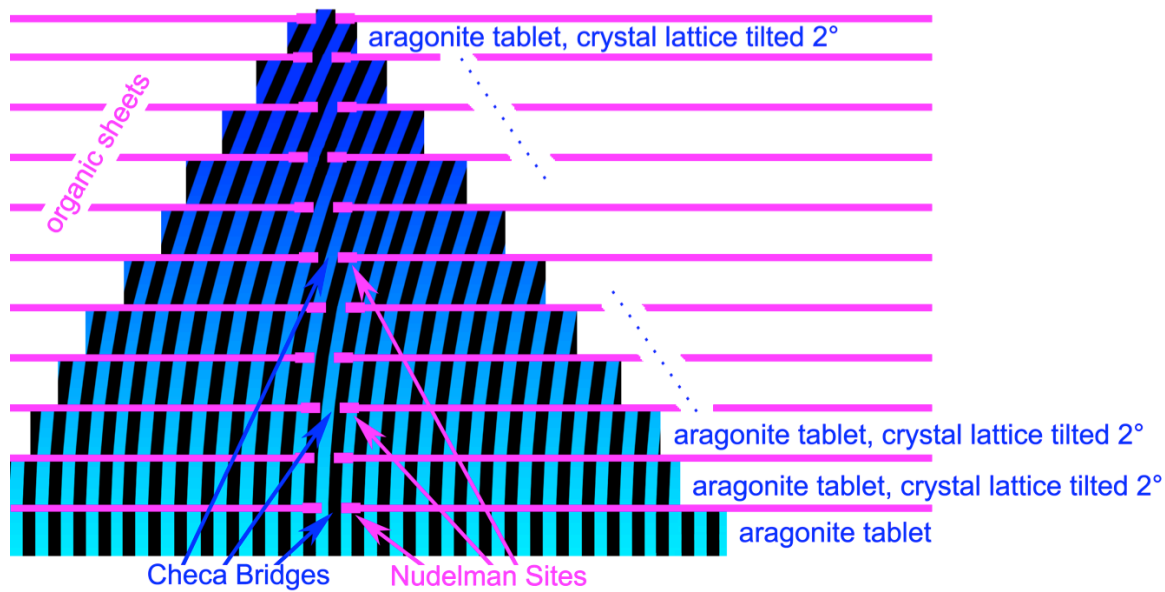

Supporting Information Figure S7. Model for nacre near-epitaxial crystal growth via "bridge tilting". Schematic representation of "bridge-tilting" in one stack of 11 tablets in columnar nacre during its growth. The bottom tablet formed first and is broader, the others have decreasing widths, will grow horizontally until they abut other tablets, and vertically they will grow epitaxially through "Checa bridges", as shown in Figure 1E. The black dashing is vertical in the bottom tablet, and then tilts in orientation by 2° at each "Checa bridge" connecting two tablets. Hence the subsequent patterns of lines are 2°, 4°, 6°, …., 20° tilted with respect to the bottom pattern, whereas the tablet surfaces remain horizontal. Growth via "bridge tilting" results in crystal-lattice tilting, and is termed "near-epitaxial" in this work.

In cases such as the dark column in Figure 5, stacks of tablets show a gradient of gray levels, rather than the more common co-oriented stacks. A possible explanation for this phenomenon assumes a single "Checa bridge", which is relatively small (~200 nm). When the "Checa bridge" protrudes as the one at the top of this schematic, it is exposed, fragile, and can easily be slightly pushed or pulled by the mantle in the living and moving mollusk. When exposed, the free-standing "Checa bridge" at the top is in direct contact with the mantle cells of the mollusk, and may be vulnerable to mantle contraction in a specific direction, resulting in bridge tilting or breaking. The subsequent tablet will grow co-oriented with its starting crystal seed, that is, the top of a "Checa bridge", in a tilted or un-tilted orientation. Thus, if there is a systematic push or pull by the mantle cells in one direction, subsequent tablets will still have their top and bottom surfaces parallel to all others and the usual layered nacre appearance will be observed, but the crystal lattice orientations of subsequent tablets will be gradually changing, resulting in the orientation gradient observed in Figure 5 and schematically represented here. This crystal lattice tilting mechanism is best explained with a single "Checa bridge" per tablet.

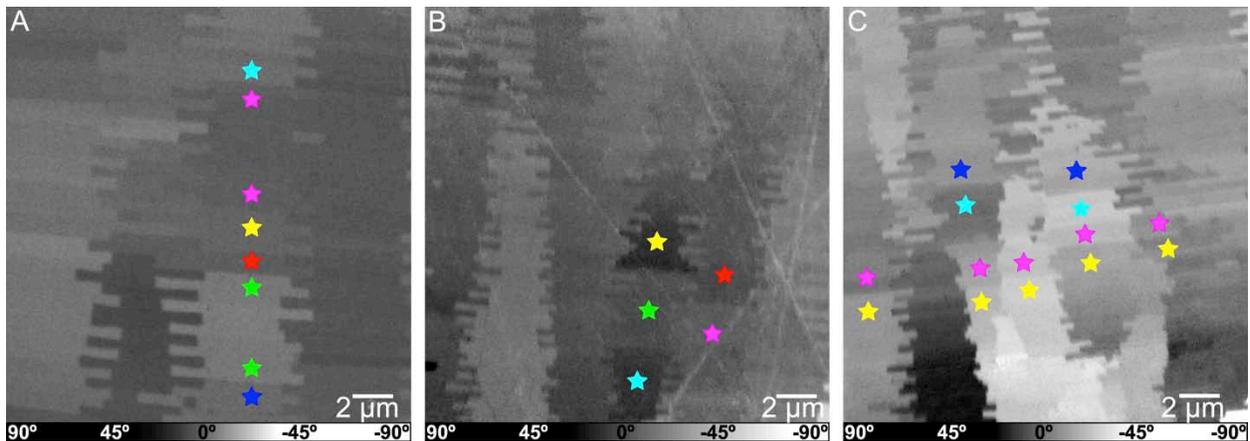

Supporting Information Figure S8. Evidence for bridge-tilting in PIC-maps of columnar nacre from 3 abalone species: (A) *Haliotis pulcherrima* (*Hp*), (B) *Haliotis discus* (*Hd*), and (C) *Haliotis rubra* (*Hrb*). In these images the nacre growth direction is from bottom to top. Most stacks of tablets (diagonal stacks in bivalve **sheet nacre**, or straight columnar stacks in gastropod or cephalopod **columnar nacre**) are composed of co-oriented aragonite crystalline tablets, as shown in most figures in this manuscript. One exception is Figure 5, where the orientation (color) in a stack varies gradually. Another exception is here, where we show abrupt changes in tablet orientations in a single stack of tablets. We note in (A) that in the stack near the center of the image, the orientation changes abruptly multiple times as nacre grows from bottom to top: the gray orientation (blue star) suddenly changes to a lighter gray level (between blue and green stars) then to a darker one (between green and red stars), lighter (yellow star), darker (magenta star), lighter again (cyan star). The abrupt crystal orientation changes occur at a specific interlamellar organic sheet. Such orientation changes, therefore, must have occurred at one "Nudelman site" or in the "Checa bridge" growing through it. (B) Similar orientation changes within one stack of tablets in *Hd* can be seen between cyan&green, and green&yellow stars, and again between magenta and red stars. (C) More sudden orientation changes are evident in five stacks of tablets, with each orientation highlighted by a colored star. We interpret all these orientation changes as the result of "bridge-tilting", because they are less than 30° in c'-axis angle spread (Supporting Information Table S1), hence slight, ±15° "bridge-tilting" is the most plausible explanation. A new nucleation event, as occurs near the NP-boundary, would have c'-axis spreads in the range between ±30° and ±90° (Figure 3). "Bridge-tilting" is therefore a more plausible explanation for the gradient of orientations shown in Figure 5.

      As an aside, an entirely speculative interpretation, which is not central to any conclusions made here or in the text, follows here. Interestingly, adjacent stacks of tablets do not always change orientations at the same nacre layer. Look for example at the sudden changes between yellow and magenta stars. The three on the left occur at the same nacre layer, but the two on the right do not. It is possible that these five changes happened

simultaneously during nacre growth, for example due to a contraction of the mantle in the mollusk, or to a change in temperature [1,2]. Not all columns grow to the same height at the same time, in fact, and different aragonite orientations grow at different crystal growth rates [3], hence simultaneously occurring events in the life of a mollusk are not necessarily recorded at the same nacre layer, but nearby. By the same reasoning, the crystal mis-orientations between cyan and blue stars may have been simultaneous. Whether they were simultaneous or not, the bridge-tilting explanation for the abrupt crystal orientation changes is unaffected.

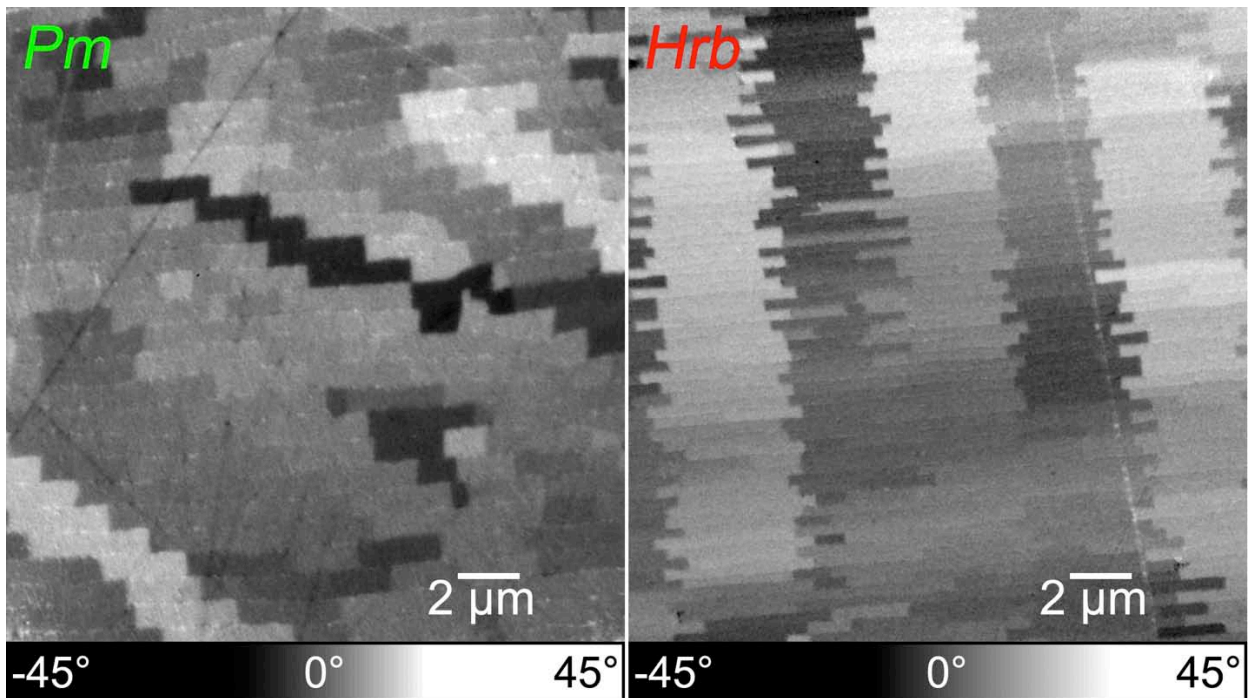

Supporting Information Figure S9. PIC-maps of nacre from *Pm* and *Hrb*. In both maps the nacre growth direction is from bottom to top. The *Pm* and *Hrb* nacre regions are 200 μm and 141 μm away from the NP-boundary, respectively, hence the decay in angle spread shown in Figures 2 and 3 has ceased, and this can be considered steady-state nacre growth. Notice that stacks of co-oriented tablets are staggered diagonally in *Pm*, and vertically in *Hrb*, as expected for sheet and columnar nacre, respectively. In *Pm*, stacks never extend more than 10 tablets with the same orientation. An abrupt change in orientation across the same geometric stack can be explained by abrupt "bridge-tilting" as shown in Supporting Information Figure S8. In *Hrb*, however, both gradual and abrupt orientation changes can be observed, as described separately in Figures 5 and Supporting Information Figure S8, respectively. Here, in each of the five vertical columns of tablets in *Hrb* nacre, both abrupt and gradual changes in crystal orientations occur. The co-existence of both kinds of changes in such close proximity strongly suggests a single mechanism as the origin:

"bridge-tilting", as described in Supporting Information Figure S7, is one possible such mechanism. Furthermore, in Figure 5 the dark column narrows from bottom to top and disappears, as the column is not parallel to the imaged two-dimensional polished surface of the sample. Here instead columns have constant width, hence bridge-tilting and column tilting do not appear to be correlated.

**Materials and Methods**

The *Atrina rigida* specimen (165 mm length), was collected at low tide after a winter storm on Sanibel Island, Florida. The *Haliotis discus* specimen (43 mm length) was farmed by the Big Island Abalone Corporation, Kona, Hawaii. The *Haliotis iris* specimen (107 mm length) from New Zealand was purchased from Australian Seashells PTY Ltd. The *Haliotis pulcherrima* specimen (55 mm length) was generously provided by Prof. Steve Weiner (Weizmann Institute of Science, Israel), collected by Prof. Heinz Lowenstam (Caltech) in Palau, Micronesia, Western Pacific Ocean, in the late 1950s or early 1960s. It was positively identified to be *Hp* by Prof. David Lindberg at UC-Berkeley. The *Haliotis laevigata* specimen (148 mm length) was provided by Prof. Monika Fritz and was originally purchased from Australian Abalone Exports PTY Ltd. (Victoria, Australia). The *Haliotis rubra* specimen (74 mm length) was farmed in Spring Bay, Tasmania. The *Haliotis rufescens* specimen (78 mm length), was farm-raised in Santa Cruz, CA and purchased from the Tokyo Fish Market in Berkeley, CA. The *Lasmigona complanata* specimen (160 mm length) and the *Pyganodon grandis* specimen (115 mm length) were collected by Prof. Sabine Gross from the Milwaukee River, and identified by Dr. Lisie Kitchel at the Wisconsin Department of Natural Resources. The *Mytilus californianus* specimen (148 mm length) was collected from the wild in Bolinas, CA. The *Mytilus edulis* specimen (54 mm length) was purchased from Hog

Island Oysters in Berkeley, CA. The *Mytilus galloprovincialis* specimen (50 mm length) from the Bay of Paimpol, France was purchased from Conchology, Inc., Philippines. *Nautilus pompilius* shells originated off the coast of Jolo Island, Philippines (141.8 mm length), and from offshore Siquijor Island, Philippines (183 mm length), and were purchased from Conchology Inc., Philippines. *Pinctada fucata* shells (58 mm length) were purchased from Hai de Ming Pearl Co. Ltd. Liusha Town, Zhanjang, China. *Pinctada margaritifera* shells (90 mm length) were purchased from the Gauguin Pearl Farm, and were farm-raised in the inner lagoon of the Rangiroa atoll, French Polynesia.

**Sample preparation:**

All shell samples were cut with a chisel and hammer. The *Nautilus pompilius* samples were taken from the outer wall of the largest chamber, not from a septum. Samples were sized with sandpaper, mounted vertically, and embedded in epoxy (either EpoThin, Buehler, IL, or EpoFix, Electron Microscopy Sciences, PA), and polished with decreasing size alumina grit down to 50 nm (MasterPrep, Buehler, IL). Sample surfaces were coated as described in refs. [4] and [5] to prevent charging phenomena [6], and still analyze the shell sample, thus the coating must be thinner than the escape depth of the photoelectrons [4]. We first deposited a silicon wafer mask at the center of the sample, where PIC-mapping analysis will be done, then coated with 40 nm of platinum using a sputter coater (208HR, Cressington, UK). We then removed the mask and slowly (~30 seconds) coated with 1 nm platinum the entire sample surface while spinning and tilting the samples [4,5].

**XANES-PEEM analysis:**

X-ray absorption near-edge structure (XANES) spectroscopy with X-ray PhotoElectron Emission spectro-Microscopy (PEEM) was performed using PEEM-3 on beamline 11.0.1, at the Advanced Light Source at Lawrence Berkeley National Laboratory in Berkeley, CA. X-ray polarization at PEEM-3 is controlled by a state-of-the-art Apple II elliptically polarizing undulator (EPU).

**PIC-mapping:**

Polarization-dependent Imaging Contrast (PIC) is an imaging modality that uses a XANES-PEEM instrument, and takes advantage of X-ray linear dichroism in carbonate crystals. Linear dichroism is the intensity variation of the carbonate carbon K-edge π* peak depending directly on the X-ray polarization angle [7-9]. A non-quantitative version of PIC-mapping was first introduced in 2007 [10], and has since been used extensively on carbonate biominerals [1-3,11-16]. Recently, Gilbert et al. proposed a new method that makes PIC-mapping semi-quantitative [15]. It cannot at present measure the position of the carbonate crystal c-axis, but its projection onto the plane containing the X-ray linear polarization vectors in XANES-PEEM, termed the c'-axis. In addition, PIC-mapping is sensitive only to the c-axis, not the orientation of the a- or b-axes [3,15] hence PIC-mapping is not fully-quantitative. This can be done for each 20-nm pixel in a typical 20 μm × 20 μm map, and yields PIC-maps in which the c'-axis orientation is displayed as gray level. This semi-quantitative PIC-mapping was employed by Olson et al. to produce 2-dimensional maps of c'-axis orientation in nacre [1,2], was recently reviewed in ref. [17] and is used here to show crystal orientation contrast in nacre, spherulitic aragonite, and prismatic calcite in mollusk shells, and also to measure the c'-axis angle spreads as a function of distance from the first nacre layer $N_o$.

Briefly, a PIC-map is produced as follows. 19 single PEEM images are acquired at the carbon K-edge π* peak energy, ~290.3eV, at 19 X-ray polarization angles (EPU°), varying from 0° (vertical) to 90° (horizontal) with 5° increments. Pixel intensity in each of these images correlates with the orientation of the crystallographic c-axis according to the curve y = a + b cos(EPU° + c) [15]. Fitting the intensity data from each pixel in the 19 images to this curve allows accurate determination of the orientation of the projection of the c-axis onto the EPU polarization plane, termed the c'-axis orientation, and creation of a 2-dimensional map of crystalline order [1]. These operations were performed automatically for each data set using software written by one of us (ICO) for use in WaveMetrics Igor Pro®, made available free of charge to any interested users [18].

**Angle spread measurement:**

Within a single quantitative PIC-map we measured the angle spread, which is the full width above the noise level of a histogram of "levels" in Igor Pro, that is, the gray level histogram which displays all pixel orientations. Measuring this "footprint" of the distribution of angles in a PIC-map is effective for representing the total spread of aragonite c'-axis angles across nacre tablets within a region. We note that the angle spread is a *maximum* spread between c'-axes orientations, and therefore any angles between the minimum and the maximum angle are included and consistently found in the data. The angle spread of c'-axes in nacre [1] is always an over-estimate compared to the angle spread of c-axes, hence angle spread measurements here are not fully-quantitative.

If many crystal orientations are present in a PIC-map, the histogram appears close to a Gaussian (Supporting Information Fig. S10A), if fewer, discrete orientations are present the histogram is highly asymmetric (Supporting Information Fig. S10B). The 2°-angle resolution for PIC-mapping is based on the observation of distinct peaks in gray level histograms of entire PIC-map images. In such histograms each peak represents a distinct crystal orientation, as shown in Supporting Information Figure S10B, with a smallest observable separation of 2°.

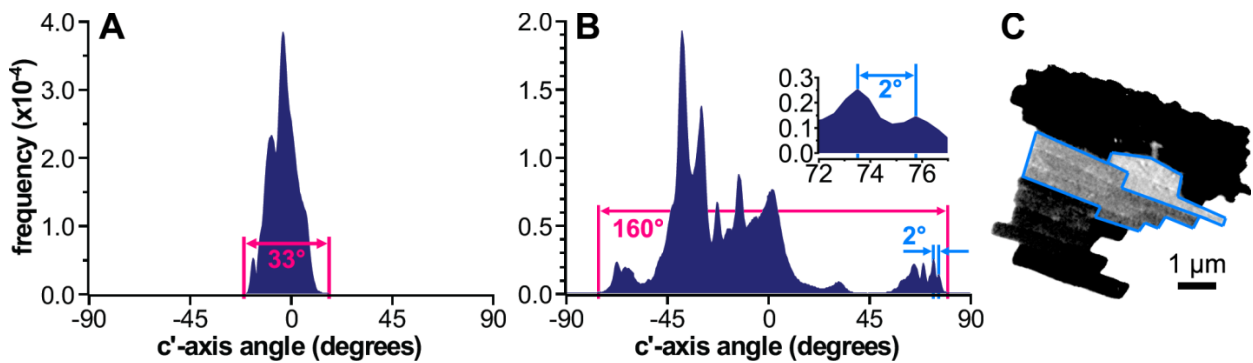

Supporting Information Figure S10. (A) Histogram of the frequency of c'-axis orientation angles from a PIC-map taken from *Lc* nacre. (B) Histogram from a region including nacre and disordered spherulitic aragonite from *HL*. The angle resolution limit of 2° is illustrated by two light blue lines, and defined as the smallest angle difference of two distinct peaks corresponding to unique tablet orientations, as shown in the inset. (C) PIC-map illustrating nacre tablets, traced in light blue, with a 2°-difference in c'-axis orientation, corresponding to the two peaks at ~74° and ~76° in B. The contrast of the PIC-map in C has been enhanced for clarity in Adobe Photoshop®. The limits in pink for measuring the full-width of the distribution in A and B were user-selected, excluding angles that were confirmed not to correspond to tablet orientations in PIC-maps as the one in C.

**Image processing:**

Partly overlapping PIC-maps in Figures 2, 4, Supporting Information Figures S1, S2, S6 were merged in Adobe Photoshop®, minor corrections to the graylevels of entire PIC-maps were made using of the gradient tool to eliminate an experimental artifact illumination gradient across a few of the images, thus minimizing contrast across adjacent images. Levels were adjusted in all merged images simultaneously to enhance contrast. The quantitative effect of leveling is indicated by the grayscale bars in these figures. Noise was removed from the PIC-maps in Figure 4 with the "Dust and Scratches" tool in Adobe® Photoshop®. The image in Figure 5 was artificially colored using the "indexed color" mode and a custom color table in Adobe Photoshop®. The clone stamp tool in Adobe Photoshop® was used to clean up unattractive few-pixel-size regions of the PIC-map in Figures 2, 4, 5, Supporting Information Figure S8.

**µXRD Analysis:**

Synchrotron Laue micro-X-ray diffraction experiments were performed on beamline 12.3.2, at the Advanced Light Source at Lawrence Berkeley National Laboratory in Berkeley, CA. The instrument uses Kirkpatrick-Baez mirror optics to focus the X-ray beam down to a size of about 1x1 µm² in cross-section at the sample position. The sample was mounted on a precision XY stage and illuminated with "pink beam" x-rays of photon energies ranging 5<hν<22 keV, at an incidence angle of 45°. X-ray microdiffraction patterns were obtained using a Pilatus 1M X-ray detector. The area detector was set to an angle of 2θ = 90° at a distance of ~140 mm from the sample. The exact detector position and orientation was calibrated using the Laue diffraction pattern of a silicon crystal. The exposure time for each diffraction pattern was 10 s.

The data of Supporting Information Figure S3 were scans with size 200 μm × 150 μm, and 2 μm step size along the nacre growth direction, and 4 μm (*Ar* and *Pm*) or 2 μm (*Hrf*) step size in the direction parallel to the NP-boundary. A Laue X-ray microdiffraction pattern from each pixel of these scans was indexed using the XMAS software (X-ray Microdiffraction Analysis) [19]. Indexing provides the full 3-dimensional orientation matrix for each crystal, allowing for the mapping of the orientations of aragonite crystallites in the sample. Analysis was performed on a 48-node Linux cluster, making it possible to automatically process thousands of Laue microdiffraction patterns, and thus map the distribution of aragonite crystal orientations in the sample.